\documentclass[12pt]{article}
\pdfoutput=1
\usepackage{amsbsy}
\usepackage{amsfonts}
\usepackage{amsmath,amssymb}
\usepackage{bbold}
\usepackage{graphicx}

\newcommand{\sect}[1]{\setcounter{equation}{0}\section{#1}}

\newcommand{\eq}{\begin{equation}}
\newcommand{\eqa}{\begin{eqnarray}}  
\newcommand{\en}{\end{equation}}
\newcommand{\ena}{\end{eqnarray}}
\newcommand{\enn}{\nonumber \end{equation}}

\def\sk{\vskip .4cm}
\def\noi{\noindent}

\def\al{\alpha}
\def\be{\beta}
\def\ga{\gamma}

\let \part\partial

\def\part{\partial}

\def\sk{\vskip .4cm}

\def\noi{\noindent}

\def\X0{X^0}

\def\al{\alpha}
\def\ga{\gamma}

\def\Dcal{{\cal D}}

\def\square{{\,\lower0.9pt\vbox{\hrule \hbox{\vrule height 0.2 cm
\hskip 0.2 cm \vrule height 0.2 cm}\hrule}\,}}

\def\Atilde{{\widetilde{A}}}

\def\phitilde{{\widetilde \phi}}

\def\lb{\langle}
\def\rb{\rangle}

\def\noali{\al_i \mkern-15mu/}

\begin{document}

\begin{titlepage}
\rightline{ARC-18-06}

\vskip 2em
\begin{center}
{\Large \bf History operators in quantum mechanics } \\[3em]

\vskip 0.5cm

{\bf
Leonardo Castellani}
\medskip

\vskip 0.5cm

{\sl Dipartimento di Scienze e Innovazione Tecnologica
\\Universit\`a del Piemonte Orientale, viale T. Michel 11, 15121 Alessandria, Italy\\ [.5em] INFN, Sezione di 
Torino, via P. Giuria 1, 10125 Torino, Italy\\ [.5em]
Arnold-Regge Center, via P. Giuria 1, 10125 Torino, Italy
}\\ [4em]
\end{center}

\begin{abstract}
\sk

It is convenient to describe a quantum system at all times by means of a ``history operator" $C$, encoding  measurements and unitary time evolution between measurements. These operators naturally arise when computing the probability
of measurement sequences, and generalize the ``sum over position histories " of the Feynman path-integral.

As we argue in the present note, this description has some computational advantages over the usual state vector description,
and may help to clarify some issues regarding nonlocality of quantum correlations and collapse.

A measurement on a system described by $C$ modifies the history operator, $C \rightarrow PC$,  where $P$ is the projector corresponding to the measurement. We refer to this modification as "history operator collapse". Thus $C$ keeps track of the succession of measurements on a system, and contains all histories compatible with the results of these measurements. The collapse modifies the history content of $C$, and therefore modifies also the past (relative to the measurement), but never in a way to violate causality. 

Probabilities of outcomes are obtained as $Tr(C^\dagger P C)/Tr(C^\dagger C)$.
A similar formula yields probabilities for intermediate measurements, and reproduces the result of the two-vector formalism in the case of  given initial and final states.

We apply the history operator formalism to a few examples: entangler circuit, Mach-Zehnder interferometer, teleportation circuit, three-box 
experiment.

Not surprisingly, the propagation of coordinate eigenstates $|q\rb$ is described by a history operator $C$ containing  the Feynman path-integral.

\end{abstract}

 \noi \hrule \vskip .2cm \noi {\small
leonardo.castellani@uniupo.it}

\end{titlepage}

\newpage
\setcounter{page}{1}

\tableofcontents

\sect{Introduction}

Can measurements affect the past ?  The question is not new, and has been raised in various forms, see for ex. \cite{retroaction}. In the perspective we adopt here,  the ``history collapse" due to a measurement indeed affects the system also at times antecedent to the measurement, since it affects its history description involving all times back to an initial state. It could seem that such a retroaction is an artifact of the history approach.  
However consider for example entangled states of two qubits that are spacelike separated. If we insist on instantaneous collapse in {\sl every} reference frame\footnote{No experimental evidence for reference-dependent ``speed of collapse" has ever been found, see for ex. \cite{speedcollapse,speedcollapse2} and references therein.}, a measurement on one qubit affects the composite system at different proper times of the second qubit, corresponding to the different reference frames where the collapse is observed. Therefore collapse must affect the whole {\it history} of the quantum system. 
This conclusion seems unavoidable if special relativity must hold together with simultaneity of collapse.

On the other hand, even nonrelativistic quantum mechanics has a spacetime description in terms of sum over histories,
i.e. Feynman path-integrals \cite{FH}. Feynman's approach has prompted various formalisms 
based on histories rather than on state vectors (for a very incomplete list of references see \cite{histories1} - \cite{histories10}). 
These formalisms must reproduce the standard probability rules of quantum mechanics, and in this sense do not add
any fundamental novelty to the theory. However they can be of help in describing closed quantum systems (containing
both the observer and the observed subsystem, as in cosmological models) and in interpreting some highly debated foundational issues.

We propose in this note to describe a quantum system by a ``sum over histories" operator, in terms of which all the rules
of standard quantum mechanics can be reproduced. This operator is just the evolution operator of the system, 
where projectors have been inserted to account for (projective) measurements. 

This history operator is more versatile than the usual state vector: it contains
in a transparent way all histories of the system, compatible with initial state and measurement results at different times.
Note that in general these histories do not form a decoherent set. This is an important difference with the {\it consistent histories} approach of ref.s \cite{histories1}-\cite{histories3} and \cite{histories5}-\cite{histories9} , where histories are required to decohere.

A further measurement produces a collapse of the history operator, by application of the projector corresponding
to the measurement result. Its ``history content" gets in general reduced, since
the projector filters out histories not compatible with the result. The conceptual consequence of this description
is that a measurement indeed affects the past, in the sense that it affects histories that go back to the initial state.

The plan of the paper is as follows. In Section 2 the familiar decomposition of the evolution operator is recalled,
and chain operators are introduced, together with the probability of obtaining sequences of measurement results.
In Section 3 the probability rules are formulated exclusively in terms of the history operator, and we discuss history amplitudes
and interference.  The collapse of the history operator is described in
Section 4, and in Section 5 we recall the probability rules for measurement results at intermediate times. These rules reproduce
the probability formula of the so called two-vector approach of \cite{twostate0}-\cite{twostate3} for a measurement  between known inititial and final states. In Section 6 the history operators of three simple quantum circuits are discussed. Section 7 deals with the
three-box experiment of \cite{threebox}. Section 8 contains some conclusions.

\sect{Evolution operator as a sum on histories}

The time evolution operator $U(t_n,t_0)$ between the times $t_0$ and $t_n$ of a quantum system can be written in a ``sum over histories" fashion as follows:
\eqa
U(t_n,t_0) &=& U(t_n,t_{n-1}) ~ U(t_{n-1},t_{n-2})  \cdots U(t_1,t_0)=~~~~~~~~~~~~~~~~~~~~~~~~~~~~~~~~ \nonumber \\
& = & U(t_n,t_{n-1})  \sum_{\alpha_{n-1}} ~ P^{(n-1)}_{\alpha_{n-1}} ~ U(t_{n-1},t_{n-2})  \cdots \sum_{\alpha_{1}} P^{(1)}_{\alpha_{1}} ~
U(t_1,t_0)  \label{evolution}
\ena
with $t_0 < t_1< \cdots < t_{n-1} < t_n$. The sum over projectors $P^{(i)}_{\alpha_{i}} $ are decompositions of the unity:
\eq
I=\sum_{\alpha_{i}} P^{(i)}_{\alpha_{i}}  \label{completeness}
\en
inserted at times $t_1,\cdots t_{n-1}$. The $P^{(i)}_{\alpha_{i}} $ are projectors on eigensubspaces of 
observables, satisfying
\eq
P^{(i)}_{\alpha_{i}} P^{(i)}_{\beta_{i}}= \delta_{\alpha_{i},\beta_{i}} P^{(i)}_{\alpha_{i}}
\en
Thus the time evolution operator is given by a sum on all indices $\al = (\al_1,\al_2,\cdots \al_{n-1})$
\eq
U(t_n,t_0) = \sum_\alpha C_\alpha
\en
where the operators $C_\alpha$ correspond to the single histories $(\al_1,\al_2,\cdots \al_{n-1})$:
\eq
C_\al =  U(t_n,t_{n-1}) ~ P^{(n-1)}_{\alpha_{n-1}} ~ U(t_{n-1},t_{n-2})  \cdots P^{(1)}_{\alpha_{1}}~
U(t_1,t_0)  \label{chain}
\en
and are usually called ``chain operators". 

A well-known exercise is to compute the
probability of obtaining a sequence of measurement results,  at times $t_1,t_2,\cdots t_{n-1}$, on a system that starts in the initial state $|\psi \rb$ at time $t_0$. If the outcomes correspond  to the projectors 
$P^{(1)}_{\al_1}, P^{(2)}_{\al_2}, \cdots P^{(n-1)}_{\al_{n-1}}$, and $P_\psi = |\psi\rb \lb \psi|$ is the projector on the initial state,
the answer is given by:
\eq
p(\psi, \al_1,\al_2,\cdots \al_{n-1}) =  Tr (C_\al P_\psi C_\al^\dagger ) \label{successive}
\en
and could be considered the ``probability of the history"  $\al = (\al_1,\al_2,\cdots \al_{n-1})$ starting
from state $|\psi \rb$.
At first sight this seems reasonable, since we can easily prove that the sum of all these probabilities gives 1:
\eq
\sum_\al  Tr ( C_\al P_\psi C_\al^\dagger) =1
\en
by using the completeness relations (\ref{completeness}) and unitarity of the $U(t_{i},t_{i-1}) $ operators.
We also find
\eq
\sum_{\al_{n}} p(\psi, \al_1,\al_2,\cdots  ,\al_{n})  =  p(\psi, \al_1, \al_2, \cdots,  \al_{n-1}) \label{sumrule0}
\en
However other standard sum rules for probabilities are not satisfied in general. For example relations of the type
\eq
\sum_{\al_2} p(\psi, \al_1,\al_2,\al_3)  =  p(\psi, \al_1, \al_3) \label{sumrule}
\en
hold only if the so-called {\it decoherence condition} is satisfied:
\eq
Tr (C_\be P_\psi C_\al^\dagger) + c.c.= 0~~when~\al \not= \be  \label{decocondition}
\en
If all the histories we consider are such that the decoherence condition holds, they are said to form a {\it consistent} set,
and can be assigned probabilities satisfying all the standard sum rules.

In general, histories do not form a consistent set: interference effects between them can be important, as in the
case of the double slit experiment. For this reason we do not limit ourselves here to consistent sets. The price
to pay is to give up the possibility of assigning a probability to each history, but this is not the goal of the 
history operator formalism. Formula (\ref{successive}) for the probability of successive measurement outcomes holds in any case, and is all we need to
compute the probabilities in terms of the history operator, as discussed in the next Sections.

\sect{History operator}

The state vector after the $\al=(\al_1,\al_2,\cdots \al_{n-1})$ measurement outcomes is obtained by applying the chain operator (\ref{chain}) to the initial state $|\psi \rb$:
\eq
| \psi_\al \rb= C_\al |\psi \rb \label{psiprime}
\en
up to a normalization factor\footnote{Due to the projectors in $C_\al$, the state $|\psi_\al \rb$ is not normalized.}. 

Suppose now that we perform an additional measurement  $P^{(n)}_{\al_n}$ on $|\psi_\al\rb$. Using the standard formula  we find:
\eq
p(\psi_\al, \al_n)= {\lb \psi_\al | P_{\al_n}^{(n)} | \psi_\al \rb \over \langle \psi_\al |\psi_\al \rangle}
\en
for the probability of obtaining the result $\al_n$. Substituting $|\psi_\al \rb$ as given in (\ref{psiprime}) yields:
\eq
p(\psi_\al, \al_n)=  p(\al_n | \psi, \al_1, \cdots \al_{n-1}) = {\lb \psi| C_\al^\dagger P_{\al_n}^{(n)} C_\al | \psi \rb \over \langle \psi| C_\al^\dagger C_\al | \psi \rangle} = 
 {Tr (C_\al^\dagger P_{\al_n}^{(n)}  C_\al P_\psi) \over  Tr ( C_\al^\dagger C_\al P_\psi )  } \label{prob1}
\en
This is the probability of obtaining the result $\al_n$ on a state that has evolved from $|\psi \rb$ to $|\psi_\al \rb$ through a sequence of measurements with results $\al_1, \cdots \al_{n-1}$. Notice that the numerator is the joint probability $p(\psi, \al_1, \cdots \al_{n-1}, \al_n)$ of obtaining  $\al_1, \cdots \al_{n-1}, \al_n$, and the denominator is the joint probability $p(\psi, \al_1, \cdots \al_{n-1})$
of obtaining  $\al_1, \cdots \al_{n-1}$. The ratio correctly gives the conditional probability of obtaining $\al_n$, if 
$\al_1, \cdots \al_{n-1}$ have been obtained.
 
 Formulae (\ref{successive}) and  (\ref{prob1}) suggest to describe the system that has evolved from $|\psi \rb$ to $|\psi_\al \rb$ via the ``history operator" 
 \eq
 C_{\psi,\al} \equiv C_\al P_\psi \label{histop}
 \en
encoding the sequence of measurements, the unitary evolution between them, and the initial state $|\psi\rb$. All relevant information on the system can be extracted from $ C_{\psi,\al}$: for example the joint probability in (\ref{successive}) can be rewritten more compactly as
\eq
p(\psi, \al_1,\al_2,\cdots \al_{n-1}) =  Tr (C_{\psi,\al}^\dagger C_{\psi,\al} ) \label{successive2}
\en
and the conditional probability 
(\ref{prob1}) takes the form
\eq
p(\al_n | \psi, \al_1, \cdots \al_{n-1}) = 
 {Tr (C_{\psi,\al}^\dagger P^{(n)}_{\al_n}  C_{\psi,\al}) \over  Tr (C_{\psi,\al}^\dagger C_{\psi,\al}) } \label{prob2}
\en
A particular history operator is the unitary evolution operator (\ref{evolution}) times $P_\psi$, corresponding to absence of measurements,
and containing all possible histories originating from the initial state $|\psi \rb$. If measurements are performed, the 
history operator contains the corresponding projectors.

It is worthwhile to emphasize, even if it is tautological, that the decompositions of unity in (\ref{evolution}), and hence the histories
contained in the evolution operator, depend on the observables measured by the experimental apparatus at the various times $t_1, \cdots , t_{n-1}$. In this way the histories contained in history operators, even in absence of actual measurements, encode information on the
measuring devices used to probe the system.

To summarize, the history operator of a system with initial state $|\psi\rb$ at $t_0$, with measuring devices that can be activated
at times $t_1, \cdots t_{n-1}$, is given by
\eq
\sum_\al C_\al P_\psi \label{historyoperator1}
\en
with the chain operators $C_\al$ given in (\ref{chain}). If actual measurements are performed at some of the times $t_i$, with results $\be_i$,
the history operator is obtained simply by replacing in  (\ref{historyoperator1}) the identity decomposition  at time $t_i$ with the single projector $P_{\be_i}^{(i)}$ .

The {\sl history content} of (\ref{historyoperator1}) is defined to be the set of all histories $\psi,\al_1, \cdots \al_{n-1}$ contained in the sum (\ref{historyoperator1}), where each history corresponds to a particular $C_\al P_\psi$.

It may be convenient to insert a decomposition of the unity also at time $t_n$, so that the history operator contains
all histories $\psi,\al_1,\cdots \al_n$ compatible with the measurement outcomes (if measurements occur) at times
$t_1, \cdots t_n$. Then the history operator becomes a sum of chain operators of the form
\eq
C_{\psi,\al}=P^{(n)}_{\al_n}~ U(t_n,t_{n-1}) ~ P^{(n-1)}_{\alpha_{n-1}} ~ U(t_{n-1},t_{n-2})  \cdots P^{(1)}_{\alpha_{1}} ~ U(t_1,t_0)  P_\psi
\label{chaincomplete}
\en
The single chain operator vanishes if $Tr (C_{\psi,\al}^\dagger C_{\psi,\al})=0$, i.e. if the joint probability of successive
measurement outcomes $p(\psi,\al_1,\cdots \al_n)$ vanishes\footnote{This probability is also called the {\sl weight} of
the history $(\psi,\al_1,\cdots \al_n)$.}. In this case the particular history 
$(\psi,\al_1,\cdots \al_n)$ is not present in the sum. 

\subsection{History amplitudes}

If the initial and final states are respectively $|\psi\rb$ and $|\phi\rb$, the history operator becomes a sum
on chain operators of the type:
\eqa
& & C_{\psi,\al,\phi}=  |\phi\rb \lb \phi | U(t_n,t_{n-1}) ~ P^{(n-1)}_{\alpha_{n-1}} ~ U(t_{n-1},t_{n-2})  \cdots P^{(1)}_{\alpha_{1}} ~ U(t_1,t_0)  |\psi\rb \lb \psi | \nonumber \\
& & ~~~~~~~~\equiv  |\phi\rb ~ A(\psi,\al,\phi) ~ \lb \psi |
\ena
The (complex) number $A(\psi,\al,\phi)$ is the {\sl amplitude} for the history $(\psi,\al,\phi)$, and it is immediate to see that
\eq
|A(\psi,\al,\phi)|^2 = Tr (C_{\psi,\al,\phi}^\dagger C_{\psi,\al,\phi})=p(\psi,\al,\phi)
\en
The history operator is then 
\eq
   ( \sum_\al  A(\psi,\al,\phi)) ~~  |\phi\rb \lb \psi |
    \en
In the sum of the amplitudes cancellations can occur when histories interfere destructively.
Computing the amplitudes for each history enables to find which histories are interfering, constructively or destructively.

\sect{Collapse}

Suppose $C_{\psi,\al}$ is the history operator of a system that has initial state $|\psi\rb$ and has been measured at times $t_1,\cdots t_{n-1}$ with outcomes $\al_1,\cdots ,\al_{n-1}$. Immediately after a measurement at time $t_n$ yielding $\al_n$, the history operator becomes
\eq
C_{\psi,\al'}=  P^{(n)}_{\al_n}  C_{\psi,\al}
\en
It describes the system that starts at $|\psi \rb$ and undergoes the $\al' = (\al_1, \cdots , \al_n)$ measurements.
The probability in (\ref{prob2}) can be expressed in terms of the history operators before and after the collapse:
\eq
p(\al_n | \psi, \al_1, \cdots \al_{n-1}) =  {Tr ({C}_{\psi,\al'}^\dagger  C_{\psi,\al'}) \over  Tr (C_{\psi,\al}^\dagger C_{\psi,\al}) } \label{prob3}
\en
 In general measurements project out some histories, but in some cases they may ``open up" histories that are
forbidden (due to interference) without the measurement. Explicit examples of ``history reduction"  and ``history restoration" will be discussed in Section 6.

The collapse of the history operator describes the modification of the history content of the system when
subjected to a measurement. Modifying history implies modifying the past, which may sound paradoxical. Note however
that no causality violation is permitted. Whether Bob measures or not the system at time $t_n$ cannot in any way
communicate information to Alice (or to himself!)  at time $t_i < t_n$. Indeed suppose for example that Alice and Bob make
measurements respectively at $t_1$ and $t_2$. The results will distribute themselves into the various values $\al_1,\al_2$ 
according to the probabilities
\eq
 p(\psi,\al_1,\al_2) = Tr(C_{\psi,\al_1,\al_2}^\dagger  C_{\psi,\al_1,\al_2})
\en
The probability for Alice to get a particular $\al_1$ is given by the sum
\eq
\sum_{\al_2} p(\psi,\al_1,\al_2)
\en
On the other hand because of property (\ref{sumrule0}) this sum is equal to $p(\psi,\al_1)$, i.e. the probability that Alice measures the particular $\al_1$ value without reference to future measurements by Bob. Thus the measuring act of Bob at $t_2$ cannot influence the statistics of measurements by Alice at times antecedent to $t_2$.

\sect{Probabilities at intermediate times}

We have so far considered two kinds of probabilities :
\sk
$\bullet$ the joint probability of a sequence of measurements, eq. (\ref{successive2}).

$\bullet$ the conditional probability of an outcome $\al_n$ at $t_n$, given the initial state $|\psi \rb$ 

~~~and the outcomes $\al_1,\cdots \al_{n-1}$
at preceding times, eq. (\ref{prob2}).
\sk
\noi We can also compute the conditional probability of measurement outcomes $\be_i$ {\it inside} the time
interval $[t_0,t_n]$, i.e. the probability that a measurement at $t_i$ yields the outcome $\be_i$, given that the
measurements at $t_0, t_1, \cdots, t_n$ ($t_i$ excluded) have as outcomes $\psi, \al_1, \cdots , \al_n$ ($\al_i$ excluded).
This probability can be expressed via the history operator $C_{\psi,\al \leftarrow \be}$ corresponding
\footnote{The notation $\al \leftarrow \beta$ is short for the sequence $\al_1, \cdots , \be_i, \cdots, \al_n$.} 
to the measurement results
$\psi, \al_1, \cdots , \be_i, \cdots, \al_n$ :
\eq
p(\be_i|\psi,\al_1,\cdots, \noali~~,\cdots \al_n) = { p(\psi,\al_1,\cdots, \be_i,\cdots \al_n)  \over \sum_{\ga_i} p(\psi,\al_1,\cdots, \ga_i,\cdots, \al_n)}
= {Tr(C_{\psi,\al \leftarrow \be}^\dagger C_{\psi,\al \leftarrow \be}) \over \sum_\ga Tr(C_{\psi,\al \leftarrow \ga}^\dagger C_{\psi,\al \leftarrow \ga})} \label{intermediate}
\en
This formula generalizes the one for the conditional probability of outcomes at time $t_n$ given in (\ref{prob3}). Note however that
\eq
 \sum_{\ga_i} p(\psi,\al_1,\cdots, \ga_i,\cdots, \al_n) \not= p(\psi,\al_1,\cdots, \noali~~,\cdots \al_n)
 \en
 \noi cf. discussion after (\ref{sumrule0}).
 
Consider the particular case when only the initial and final states of the system
 are given, respectively as $|\psi \rb$  at $t_0$ and $|\phi \rb$ at $t_2$ . Then formula (\ref{intermediate}) yields
 \eq
 p(\beta|\psi,\phi)={Tr(C^\dagger_{\psi,\beta,\phi} C_{\psi,\beta,\phi} ) \over \sum_\ga Tr(C^\dagger_{\psi,\ga,\phi} C_{\psi,\ga,\phi} )}
 \label{twovector}
 \en
 for the probability of obtaining the result $\be$ in a measurement at time $t_1$. The history operator is in this case:
 \eq
 C_{\psi,\beta,\phi} = P_\phi U(t_2,t_1) P^{(1)}_\be U(t_1,t_0) P_\psi
 \en
 so that
 \eqa
  & & Tr(C^\dagger_{\psi,\beta,\phi} C_{\psi,\beta,\phi} ) =Tr(U(t_1,t_0) P_\psi U^\dagger (t_1,t_0) P^{(1)}_\be U^\dagger (t_2,t_1) P_\phi  U(t_2,t_1) P^{(1)}_\be ) \nonumber \\
  & & ~~~~~~~~~~~~~~~~~~~~~= |\lb \psi(t_1)| P^{(1)}_\be |\phi (t_1) \rb|^2
  \ena
 using $|\psi(t_1) \rb = U(t_1,t_0) |\psi \rb$ and $|\phi(t_1) \rb = U(t_1,t_2) |\phi \rb$. Therefore (\ref{twovector}) becomes
 \eq
  p(\beta|\psi,\phi)={ |\lb \psi(t_1)| P^{(1)}_\be |\phi (t_1) \rb|^2 \over \sum_\ga  |\lb \psi(t_1)| P^{(1)}_\ga |\phi (t_1) \rb|^2}
  \en
 and reproduces the symmetric formula of the two-vector formalism of \cite{twostate0} - \cite{twostate3}.
 
\sect{Examples}

In this Section we examine three examples of quantum systems evolving from a given initial state, and subjected to successive measurements.
These examples are taken from simple quantum computation circuits\footnote{A review on quantum computation can be found for ex. in \cite{QC} .} where unitary gates determine the evolution between measurements. 
Only two gates will be used: the Hadamard one-qubit gate $H$ defined by:
\eq
H |0\rb = {1\over \sqrt{2}} (|0\rb + |1\rb),~~~H |1\rb = {1\over \sqrt{2}} (|0\rb - |1\rb)
\en
and the two-qubit $CNOT$ gate:
\eq
CNOT |00\rb = |00\rb,~CNOT |01\rb = |01\rb,~CNOT |10\rb = |11\rb,~CNOT |11\rb = |10\rb
\en
Quantum computing circuits in the consistent history formalism have been discussed for example in 
in ref.s \cite{histories2,quantumcomputation1}.

\subsection{The entangler circuit}

This basic circuit uses a Hadamard gate and a $CNOT$ gate, and delivers entangled two-qubit states (the Bell states). 
Traditionally the upper qubit belongs to Alice, the lower one to Bob.

\includegraphics[scale=0.35]{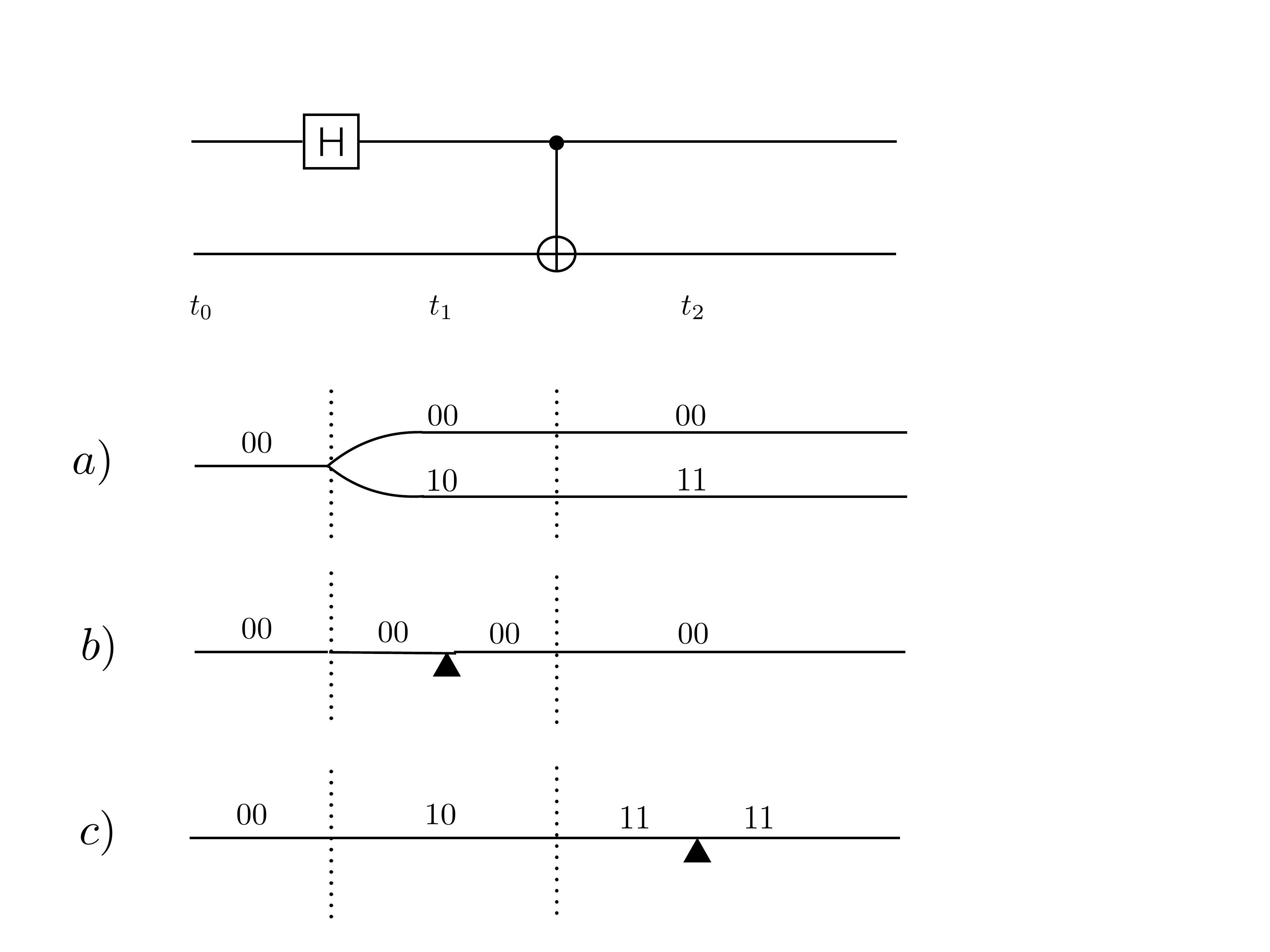}

\noi {\bf Fig. 1}  {\small The entangler circuit, and some history diagrams: a) no measurements, or Bob measures 0 at $t_1$;  b) Alice measures 0 at $t_1$; c) Alice measures 1 at $t_2$. Black triangles indicate measurements.}
\sk

\noi Consider for example the history operator
describing the system with initial state $|\psi\rb = |00\rb$. The output of the circuit is
the Bell state 
\eq
|\be_{00}\rb ={1 \over \sqrt{2}} (|00\rb + |11\rb )
\en
Before any measurement, the 
history operator is:
\eq
C =  {\rm CNOT} (H\otimes I) ~ |00\rb \lb 00| =  \sum_{\al_1,\al_2} P_{\al_2}  {\rm CNOT}~ P_{\al_1}  (H\otimes I)~ |00\rb \lb 00|
\en
and is easily seen to contain the two histories $(\psi,\al_1,\al_2)$:
\eqa
& & 00 \rightarrow 00 \rightarrow 00 \label{historiesentangler1}\\
& & 00 \rightarrow 10 \rightarrow 11 \label{historiesentangler2}
\ena
corresponding to Fig. 1 a). 

Let us now introduce measurements. For example suppose a measurement at $t_2$ has given $\al_2=11$.
The history operator becomes:
\eq
C' =  \sum_{\al_1} |11\rb \lb 11|~  {\rm CNOT}~  P_{\al_1}   (H\otimes I)~ |00\rb \lb 00|
\en
and contains only the second history (\ref{historiesentangler2}). The same happens if the measurement at
$t_2$ is performed only by Alice (then the projector at $t_2$ is $|1\rb \lb 1| \otimes I$).
An interesting consequence of this history collapse is that it involves also the past
relative to $t_2$. After Alice has obtained $1$ at time $t_2$,  the system is described by a history operator containing the single
history $00 \rightarrow 10 \rightarrow 11$. Thus a measurement by Alice on her qubit does not
``instantaneously" affect Bob's measurement statistics, but affects the {\it whole history} of the system.
 Indeed, due to history collapse the state of Bob's qubit is
$|1\rb$ {\sl even before the measurement} (see Fig. 1c), and no ``spooky action at a distance" is required.

If Bob measures his qubit at $t_1$ and obtains 0, the history diagram a) remains unchanged, since both histories
``pass through" the projector $P_{\al_1} = I \otimes |0\rb \lb 0|$. On the other hand, if Alice measures 0 at $t_1$,
only one history survives, see Fig. 1 b).

By means of formula (\ref{prob3}) one finds the probability to measure 11 at time $t_2$:
\eq
 {Tr ({C'}^\dagger  C') \over  Tr (C^\dagger C) }=  {1 \over 2}
 \en
 and similar for the probability to find 00. Using the same formula, but with
 the projector at $t_2$ given by $|1\rb \lb 1| \otimes I$, yields the probabilty (=${1\over 2}$) for Alice to measure 1 on her qubit at 
 time $t_2$.
 \sk
\noi  {\bf Note:} the two histories of the unmeasured system are orthogonal, in the sense that
 $Tr(C^\dagger_{00,00,00} C_{00,01,11})=0$,
and form therefore a consistent set\footnote{The decoherence condition (\ref{decocondition}) takes the form $Tr(C_{\psi,\al}^\dagger
 C_{\psi,\be}) + c.c =0$ in terms of history operators. When the scalar product between history operators $Tr(C_{\psi,\al}^\dagger
 C_{\psi,\be})$ vanishes, the decoherence condition is satisfied ``a fortiori".}.

\subsection{Mach-Zehnder interferometer}

The circuit  in Fig. 2 mimics a particular setting of the Mach-Zehnder interferometer, which provides a convenient discretization of 
the double slit experiment. It is a one-qubit line with two Hadamard gates:

\includegraphics[scale=0.45]{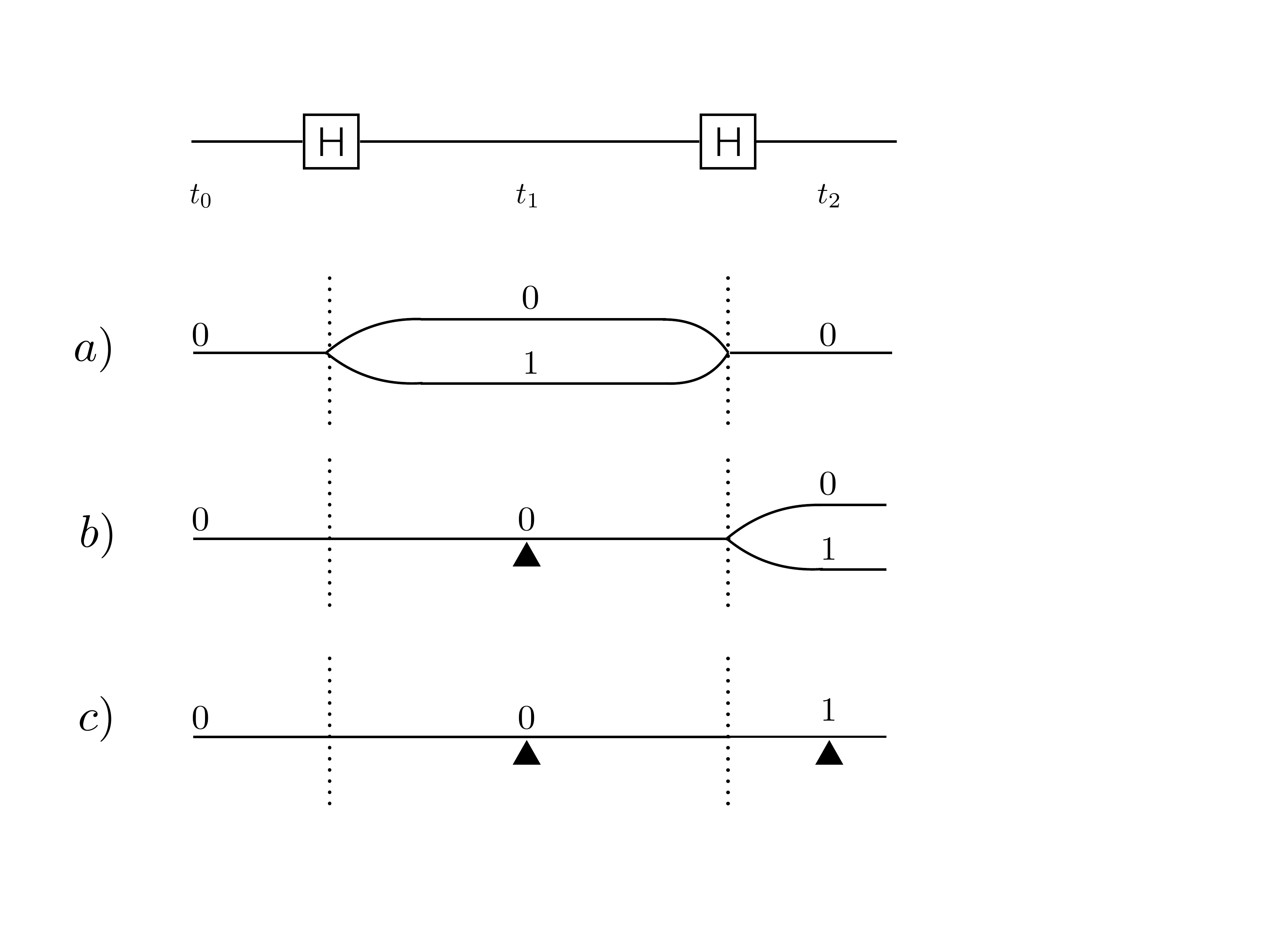}
\noi {\bf Fig. 2} {\small Circuit analogue of the Mach-Zehnder interferometer and some history diagrams with initial state $|0\rb$: a) no measurements, b) measurement giving 0 at time $t_1$, c) two measurements, giving 0 at time $t_1$ and 1 at time $t_2$.}

\sk
\noi The insertion of unity occurs at times $t_1$ and $t_2$. For an initial state $|\psi\rb$ = $|0\rb$, the history
operator of the unmeasured system is\footnote{Recall $H^2=I$.}:
\eq
C=  |0\rb \lb 0| = \sum_{\al_1,\al_2}   P_{\al_2} H~P_{\al_1} ~  H ~ |0\rb \lb 0|
\en
and contains only the two histories:
\eqa
& & 0 \rightarrow 0 \rightarrow 0  \label{historiesmachzender1}\\
& & 0 \rightarrow 1 \rightarrow 0  \label{historiesmachzender2}
\ena
while the other two histories, i.e.
\eqa
& & 0 \rightarrow 0 \rightarrow 1  \label{historiesmachzender3}\\
& & 0 \rightarrow 1 \rightarrow 1  \label{historiesmachzender4}
\ena
give opposite contributions to the history operator, and therefore disappear from the sum.
The diagram of the unmeasured history operator is given in Fig.2a: in this situation 
a measurement at $t_2$ can only have outcome 0. The histories with outcome 1 at $t_2$ 
have interfered and are not present in the history operator. The situation changes drastically
if a measurement occurs at $t_1$. Supposing that its outcome is 0, the history
operator becomes the one represented in Fig.2b, where two different histories 
open up after the second Hadamard gate, so that a further measurement at $t_2$ can
give both results, 0 or 1. In Fig.2c a further measurement at $t_2$ with outcome 1 has been added,
and the only history surviving is $0 \rightarrow 0 \rightarrow 1$, a history that was absent without the measurement in $t_1$
because of destructive interference with $0 \rightarrow 1 \rightarrow 1$.
In the double slit experiment, the measure at $t_1$ corresponds to detect the photon at one of the two slits, thereby
destroying the interference pattern on the screen (where the photon arrives at $t_2$). 
\sk
\noi {\bf Note:} the two histories of the ``unmeasured" system are {\it not} orthogonal, since
 \eq
 Tr(C^\dagger_{000} C_{010})= {1\over 2}
 \en
 and therefore are not a consistent set. Indeed interference occurs, constructive for the two histories (\ref{historiesmachzender1}), (\ref{historiesmachzender2}) and
destructive for the two histories (\ref{historiesmachzender3}), (\ref{historiesmachzender4}).
\sk
Probabilities of measurement results at times $t_1$, $t_2$ or at both times can be computed 
easily with the general formulae (\ref{prob2}) and  (\ref{prob3})  and give
\eqa
& & p(\al_1=0)=1/2,~~~p(\al_1=1)=1/2 \nonumber  \\
& &  p(\al_2=0)=1,~~~p(\al_2=1)=0 \nonumber \\
& &  p(\al_1=0, \al_2=0)=1/4,~~~p(\al_1=0,\al_2=1)=1/4,\nonumber \\
& & p(\al_1=1,\al_2=0)=1/4 ,~~~p(\al_1=1,\al_2=1)=1/4
\ena
and we can verify that:
\eq
p(\al_1=0, \al_2=0)+p(\al_1=1, \al_2=0) \not= p(\al_2=0)
\en
In this case the histories $(\psi,\al_1,\al_2)$ cannot be assigned individual probabilities.

\vfill\eject

\subsection{Teleportation circuit }

The teleportation circuit \cite{teleportation}  is the three-qubit circuit given in Fig. 3, where the upper two qubits belong to Alice, and the 
lower one to Bob.

\includegraphics[scale=0.45]{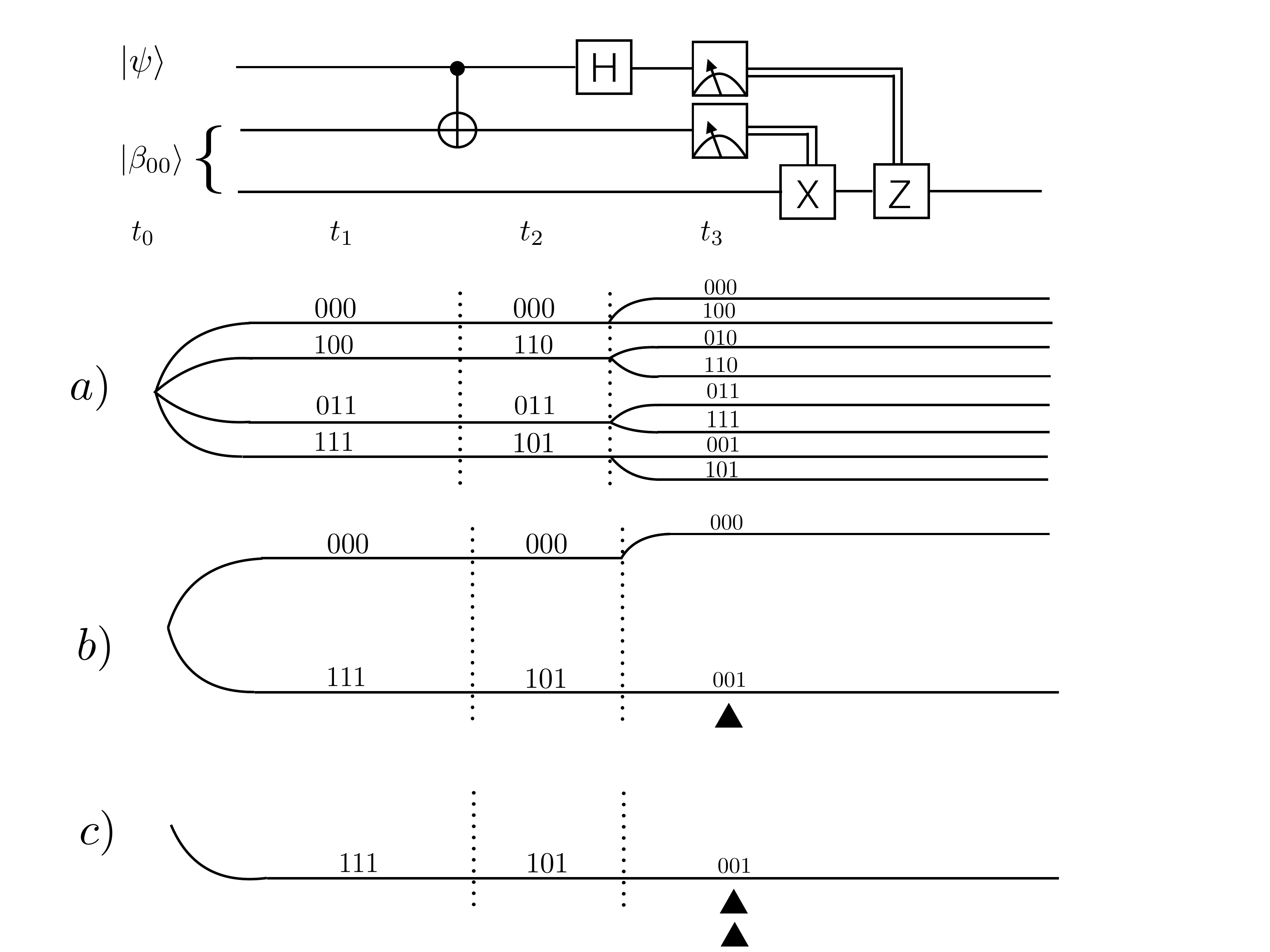}
\noi {\bf Fig. 3 } {\small Teleportation circuit: a) no measurements; b) Alice measures 00 at time $t_3$; c) at time $t_3$ Alice measures 00 
and Bob measures 1.}
\sk
\noi The initial state is a three-qubit state, tensor product of the single qubit $|\psi\rb$ to be teleported and the 2-qubit entangled Bell state
$|\be_{00}\rb = {1 \over \sqrt{2}} (|00 \rb + |11 \rb$. Before any measurement, the history operator is
\eqa
& & C =  H_1 ~ {\rm CNOT_{1,2}} ~ |\psi\rb \otimes |\be_{00}\rb   \lb \psi | \otimes \lb \be_{00} |= \\
& & ~~ =  \sum_{\al_1,\al_2,\al_3} P_{\al_3} H_1~P_{\al_2} {\rm CNOT_{1,2}} ~ P_{\al_1}  |\psi\rb \otimes |\be_{00}\rb  \lb \psi | \otimes \lb \be_{00} |
\ena
where unity as sum of projectors has been inserted at times $t_1,t_2,t_3$, and with obvious notations $H_1 \equiv H \otimes I \otimes I$ and $ {\rm CNOT_{1,2}} \equiv {\rm CNOT} \otimes I$.  For the moment we do not take into account the $X$ and $Z$
gates, activated by the results of Alice measurements at $t_3$. The history operator has the representation given in Fig. 3a
and contains 8 histories:
\eqa
& & 000 \rightarrow 000 \rightarrow 000,~~~~~000 \rightarrow 000 \rightarrow 100 \nonumber \\
& & 100 \rightarrow 110 \rightarrow 010,~~~~~000 \rightarrow 110 \rightarrow 110 \nonumber \\
& & 011 \rightarrow 011 \rightarrow 011,~~~~~011 \rightarrow 011 \rightarrow 111 \nonumber \\
& & 111 \rightarrow 101 \rightarrow 001,~~~~~111 \rightarrow 101 \rightarrow 101 
\ena

Suppose now that Alice measures her two qubits. There are four possible outcomes, each with
probability $1/4$. This can be checked by use of formula (\ref{prob3}). For example, if Alice obtains $00$,
the collapsed history operator is
\eq
C' = (|00\rb \lb 00| \otimes I)~\sum_{\al_1,\al_2}  H_1~P_{\al_2} {\rm CNOT_{1,2}}  P_{\al_1} ( |\psi\rb \otimes |\be_{00}\rb) ( \lb \psi | \otimes \lb \be_{00} |)
\en
and contains only the two histories
\eqa
& & 000 \rightarrow 000 \rightarrow 000,  \nonumber\\
& & 111 \rightarrow 101 \rightarrow 001  \label{telehistories}
\ena
\noi see Fig. 3b.
If on this system Bob measures his qubit, and obtains 0 or 1, the history operator becomes 
\eq
C''= (I_{2 \times 2} \otimes |0\rb \lb 0|)~ C' ~~{\rm or}~~C'' = (I_{2 \times 2} \otimes |1\rb \lb 1|)~ C'
\en
and contains only one of the two histories in (\ref{telehistories}), see Fig. 3c for the outcome 1.
If  $|\psi\rb$ is given by
\eq
|\psi\rb = \alpha |0\rb + \be |1 \rb
\en
formula (\ref{prob3}) yields the probabilities $|\al|^2$ and $|\be|^2$ for Bob to obtain  0 or 1 respectively, thus showing
that the state $|\psi\rb$ has been correctly teleported. 
Similar arguments hold if Alice obtains $01$ or $10$ or $11$. In these cases the gates $X$ and $Z$, represented by the Pauli
matrices $\sigma_x$ and $\sigma_z$ on the ($|0\rb$, $|1\rb$) basis, have to be added to
the history operator.

Finally, if Bob measures his qubit without any preceding measurement by Alice, the surviving histories are the 
upper four (when Bob finds 0) or the lower four (when Bob finds 1) in Fig. 3a.

\subsection{Path-integral operator}

Consider a generic quantum system with initial state $|q_0\rb$ and final state $|q_n\rb$, controlled by a Hamiltonian $H$.
Subdividing time between $t_0$ and $t_n$ as usual $t_0 < t_1 < t_2 \cdots <t_n$, and inserting unity as sum of projectors
on coordinate eigenvectors $|q_i \rb$:
\eq
I = \int dq_i |q_i \rb \lb q_i|
\en
yields the history operator of the system in the form:
\eq
C_{q_0,q_n} = \int dq_1 \cdots dq_{n-1}  |q_n\rb \lb q_n| e^{-{i\over \hbar}  H (t_n - t_{n-1})}|q_{n-1}\rb \lb q_{n-1}|  \cdots |q_1\rb \lb q_1|
e^{-{i\over \hbar}  H (t_1 - t_{0})} |q_0\rb \lb q_0|
\en
Taking the limit of infinitesimal time intervals we recover the familiar path-integral operator
\eq
C_{q_0,q_n}= |q_n\rb \lb q_0| \int \Dcal q ~ e^{-{i\over \hbar}  S (q_0,q_n)}
\en
where 
\eq
S (q_0,q_n) = \int_{q_0,t_0}^{q_n,t_n} L (q,{\dot q})~ dt
\en
and $L=p {\dot q} - H$ is the Lagrangian.
Formulae (\ref{prob2}) and (\ref{prob3}) reproduce the probability of propagation from $|q_0\rb$ to $|q_n\rb$.
In fact the multiple insertion of identity as sum over projectors on coordinate eigenvectors is the central
idea of Feynman's path-integral, an idea that has inspired all  history-based formulations of
quantum mechanics.

\sect{The three-box experiment}

It is sometimes called ``the three-box paradox", although no paradox is involved\footnote{We thank R. B. Griffiths
for calling our attention to this Gedanken experiment, and his treatment of it in ref. \cite{histories2}.}. It was introduced in ref. \cite{threebox}, and discussed in many subsequent works.

The ingredients are a (quantum) particle, three boxes $A$, $B$ and $C$, and measuring devices that probe the system
at times $t_1$ and $t_2$.  
The particle is prepared in the initial state at $t_0$:
\eq
|\psi \rb = {1 \over \sqrt{3}} (|A\rb+|B\rb+|C\rb)
\en
\noi where the mutually orthogonal states $|A\rb$, $|B\rb$ and $|C\rb$ correspond to the particle being in
box $A$, $B$, and $C$.  
\sk
Consider first the case of a measuring apparatus able to detect whether the particle
is in $A$ at $t_1$, and in the state
\eq
|\phi\rb = {1 \over \sqrt{3}} (|A\rb+|B\rb-|C\rb)
\en
at $t_2$. The relevant projectors are therefore
\eqa
& & at~t_0:~~~~~P_\psi = |\psi\rb \lb \psi|\\
& & at~t_1:~~~~~P_A = |A\rb \lb A|,~~~P_{\Atilde}=I- |A\rb \lb A| \\
& & at~t_2:~~~~~P_\phi = |\phi\rb \lb \phi|,~~~~P_{\phitilde}=I- |\phi\rb \lb \phi| 
\ena
and the history operator of the system, before any measurement, is written as
\eq
(P_\phi+P_{\phitilde})~  (P_A +P_{\Atilde})~ P_\psi
\en
By inspection it contains the histories:
\eq
\psi A \phi,~\psi A \phitilde,~\psi \Atilde \phitilde
\en
which form a consistent set. The particular history $\psi \Atilde \phi$ is absent because
\eq
P_\phi  P_{\Atilde} P_\psi = 0
\en
We can now compute the conditional probability of finding the particle in box $A$, given
a final state $|\phi\rb$. Using formula (\ref{twovector}) one finds:
\eq
p(A|\psi,\phi) = 1 \label{pA}
\en
In other words, if a particle detector is turned on in box $A$, it will always detect the particle in $A$ if the
final state of the system is $|\phi\rb$.
\sk
The so-called paradox arises when one replaces $A$ with $B$ in the above reasoning. Since $A$ and $B$ play symmetrical roles,
one finds that the conditional probability of finding the particle in box $B$, given
a final state $|\phi\rb$, is
\eq
p(B|\psi,\phi) = 1
\en
which seems to contradict (\ref{pA}) since it means that the particle will always be found in $B$, if the same final state $|\phi\rb$   
is post-selected. 

In fact no contradiction arises, because the two situations are different. In the second case the
measuring device at $t_1$ is a particle detector in box $B$, differing from the particle detector in box $A$.
One will always find the particle in $A$ if  ``looking" in the box  $A$, and always in $B$ if ``looking" in the box $B$, provided of course that the final state is postselected to be $|\phi\rb$.  This may sound counter-intuitive using classical logic, but
such ``paradoxes" are the trademark of the quantum world.

We may ask what happens if at $t_1$ the measuring apparatus is able to detect {\sl in which box} the particle sits. 
We must then use the decomposition of unity
\eq
I = P_A + P_B + P_C
\en
at $t_1$, instead of $I=P_A + P_\Atilde$. This gives rise to a different set of histories:
\eq
\psi A \phi,~\psi A \phitilde,~\psi B \phi,~\psi B \phitilde,~\psi C \phi,~\psi C \phitilde,
\en
a {\sl non-consistent} set: for example the two histories
$\psi A \phi$ and $\psi B \phi$ are not orthogonal. Nonetheless the history operator contains all of them, and
in terms of it we can compute conditional probabilities, finding:
\eq
p(A|\psi,\phi)={1 \over 3},~~~p(B|\psi,\phi)={1 \over 3},~~~p(C|\psi,\phi)={1 \over 3}
\en
So if we look in each box, again with final state $|\phi\rb$ postselected, we find the particle in box $A$,$B$ and $C$ with equal probability.

\sect{Conclusions}

We have proposed in this note to describe quantum systems by means of a history operator, that allows to compute  
probabilities in agreement with the usual state vector formulation.
There is nothing fundamentally new in this formalism, but it does provide a more immediate and diagrammatic way 
to represent the ``history content" of a quantum system. In particular, by shifting the point of view from state vectors
to histories, it helps to alleviate a conflict between
simultaneity of collapse and special relativity. 

\section*{Acknowledgements}

We thank R. B. Griffiths for interesting comments and suggestions on the first version of this work.

\vfill\eject
\end{document}